\documentclass{nature}
\usepackage{upgreek}
\usepackage{amsmath}
\usepackage{graphicx}
\usepackage[font=bf]{caption}
\usepackage{amssymb}
\title{Waveguide Integrated Superconducting Single Photon Detectors Implemented as Coherent Perfect Absorbers}

\author{Mohsen K. Akhlaghi$^{1,*}$, Ellen Schelew$^{1}$ \& Jeff F. Young$^{1,+}$ \\
$^1$\textit{Department of Physics and Astronomy, University of British Columbia}\\
\textit{Vancouver BC V6T 1Z1 Canada} \\
$^{*}$email: akhlaghi@phas.ubc.ca, $^{+}$email: young@phas.ubc.ca
}

\begin{document}

\maketitle

\begin{abstract}
At the core of an ideal single photon detector is an active material that ideally absorbs and converts photons to discriminable electronic signals. A large active material volume favours high-efficiency absorption, but often at the expense of  conversion efficiency, noise, speed and timing accuracy. The present work demonstrates how the concept of coherent perfect absorption can be used to relax this trade-off for a waveguide-integrated superconducting nanowire single photon detector. A very short (8.5\(\upmu\)m long) and narrow (8$\times$35nm$^2$) U-shaped NbTiN nanowire atop a silicon-on-insulator waveguide is turned into a perfect absorber by etching an asymmetric nanobeam cavity around it. At 2.05K, the detectors show $\sim$96$\pm$12\% on-chip quantum efficiency for 1545nm photons with an intrinsic dark count rate $<$0.1Hz. The estimated timing jitter is $\sim$53ps full-width at half-maximum and the reset time is $<$7ns, both extrinsically limited by readout electronics. This  architecture is capable of pushing ultra-compact detector performance to ideal limits, and so promises to find a myriad of applications in quantum optics.
\end{abstract}

\section{Introduction}

Over the past decade, photonic quantum information processing\cite{Bennett1995,OBrien2009} has shown considerable promise for quantum computing\cite{OBrien2007,Kok2007,Knill2001}, communication\cite{Gisin2007} and metrology\cite{Giovannetti2006,Mitchell2004,Nagata2007}. Most of the impressive proof-of-principle demonstrations have relied on bulk optics\cite{OBrien2009} that is inherently non scalable\cite{Tanzilli2012}. Independent progress on conventional integrated photonic circuitry\cite{Smit2012,Streshinsky2013,Hunsperger2009,Tien1977} suggests that integrated quantum photonic circuits\cite{Tanzilli2012} may offer a scalable solution. A variety of pertinent host material systems\cite{Politi2008,Marshall2009,Peruzzo2010,Bonneau2012,Zhang2011,Wang2014,Greentree2008} are being studied, but the unparalleled recent progress in classical silicon-based photonic circuits\cite{Baehr-Jones2012,Streshinsky2013,Reed2005,Xia2011} makes silicon particularly attractive. Phase modulators and a wide assortment of ultra-compact (tunable) passive components have been demonstrated in the silicon-on-insulator (SOI) platform\cite{Lipson2005}.  Already some of these have been integrated\cite{Takesue2013,Silverstone2014} with on-chip entangled and heralded photon sources\cite{Sharping2006,Davanco2012,Takesue2007,Takesue2008,Luxmoore2013,Azzini2012,Xiong2011}, single photon detectors\cite{Pernice2012}, and used to demonstrate quantum interference and manipulation\cite{Bonneau2012a,Xu2013}.

All the optical components on a SOI platform - including the detectors - should operate at telecom compatible infrared wavelengths. Superconducting nanowire single photon detectors (SNSPD)\cite{Hadfield2009, Natarajan2012} represent the most promising stand-alone infrared photon counting technology, and so it is not surprising that nanowires placed on top of optical waveguides have figured prominently in recent demonstrations of integrated single photon detectors\cite{Sahin2013,Sprengers2011,Schuck2013,Atikian2014,Calkins2013,Pernice2012}.  In the travelling wave (TW) configuration employed in these detectors, photons propagating down micron-wide waveguides are evanescently absorbed by a critically biased superconducting nanowire over a distance of tens of microns, and then an easily detectable normal state transition in the nanowire occurs\cite{Natarajan2012}. Various of these TW SNSPDs have achieved high quantum efficiencies  (up to 91\%\cite{Pernice2012}), or low noise (down to $<$0.01Hz\cite{Schuck2013a}), or fast recovery time ($<$10ns\cite{Pernice2012}), or accurate timing response ($<$20ps of jitter\cite{Pernice2012}), but  no integrated single detector performs well in all of these critical categories. If truly scalable quantum information on-a-chip is to become practical, then  detectors that exceed all of these individual performance metrics are needed.

\begin{figure}[h!]
\centering
\includegraphics[width=6.2in]{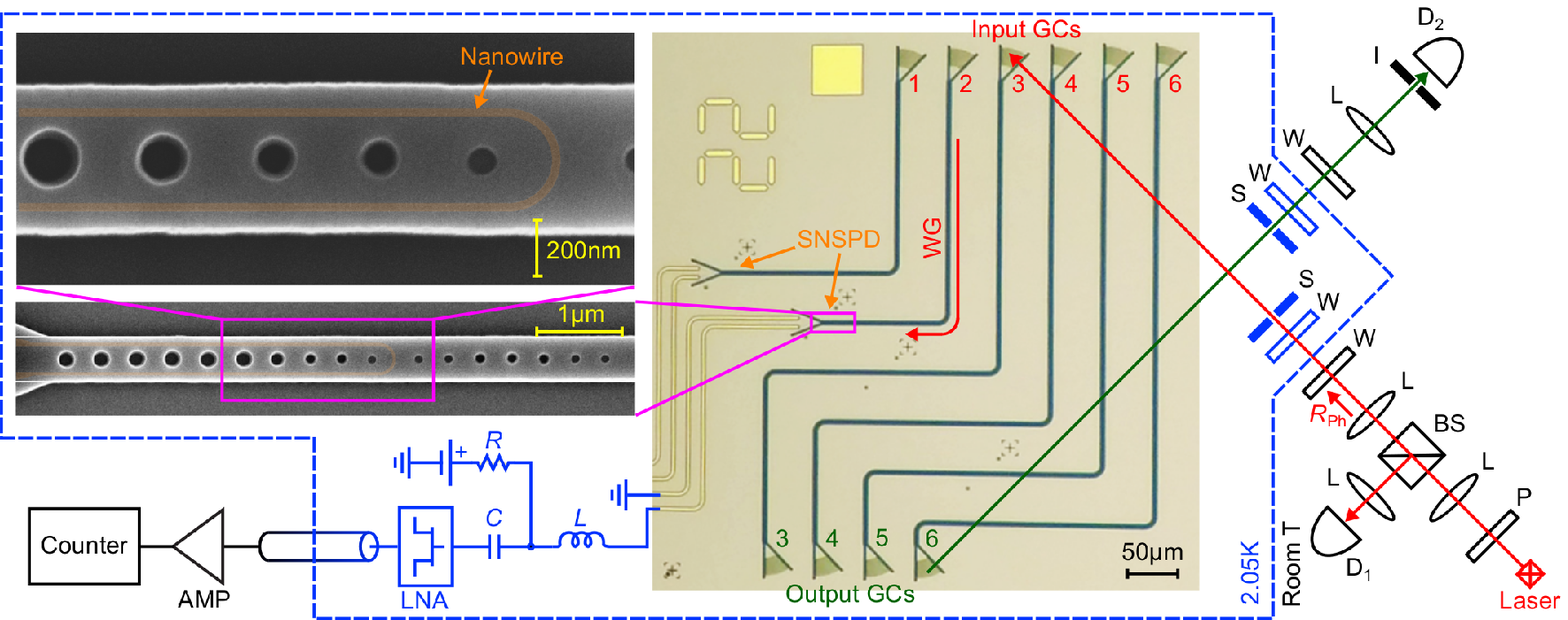}
\caption{A waveguide integrated SNSPD designed as a coherent perfect absorber by patterning an asymmetric nanobeam cavity around a superconducting nanowire. \textmd{The optical micrograph at the center shows a typical chip that includes two SNSPDs connected to grating couplers (GC) number 1 and 2 via two waveguides (WG). The other devices are for calibration purposes. Insets on the left are scanning electron microscope images (nanowire is colored) zoomed into the detector region. The dashed blue line encloses components held at cryogenic temperature (2.05K). The photons generated by a laser are delivered to the detector through waveguides, grating couplers, a hole in a cold shield surrounding the chip (S) and room-temperature optics (P: polarizer, L: lens, BS: beam-splitter, W: cryostat windows, $\text{D}_{1}$: power meter). GC-WG-GC devices together with a second set of room temperature optics that involves $\text{D}_{2}$ are used for calibration, imaging and alignment purposes (I: iris). The nanowire is biased by a voltage source, a resistor ($R=100$\(\Omega\)), and an inductor ($L=100\text{nH}$). The detection signal is devlivered to a low noise amplifier (LNA) through a coupling capacitor ($C=22\text{pF}$) for transmission to a room temperature amplifier (AMP) and a counter through a coaxial cable.}}
\label{fig:1}
\end{figure}

The evanescent absorption in the TW geometry intrinsically puts a minimum on the coupling length of an efficient detector, and correspondingly, the length of the nanowire (typically between 40\(\upmu\)m and 400\(\upmu\)m \cite{Pernice2012,Sprengers2011,Schuck2013} depending on the nanowire layout and the host material system).  A higher packing density of lower noise\cite{Bulaevskii2011}, faster\cite{Kerman2006,Kerman2009} and more accurate detectors\cite{OConnor2011} would be possible if the nanowire length could be further reduced.  The obvious challenge is the apparent trade-off in absorption efficiency\cite{Kovalyuk2013} (shorter wires offer less absorbing volume).  This paper describes a successful strategy for circumventing this trade-off, by embedding a short (8.5\(\upmu\)m) ultra-narrow (8$\times$35nm$^2$) superconducting nanowire within a high quality factor microcavity specifically designed to turn the overall detector into a coherent perfect absorber\cite{Wan2011,Chong2010,Zhang2012}  (see scanning electron microscope images on Fig.~\ref{fig:1}).  Implemented on a SOI platform, the detector's ultra small foot-print (0.5$\times$7.0\(\upmu\)m$^2$) - smaller than any reported to date - also incorporates a built-in optical filter. It absorbs and detects nearly 100\% of $\sim$1545nm light in the waveguide while exhibiting a sub-Hz intrinsic dark count rate, and a fast recovery time $<$7ns.

\section{Results}

\subsection{Design Concept}

\begin{figure}[h!]
\centering
\includegraphics[width=6.2in]{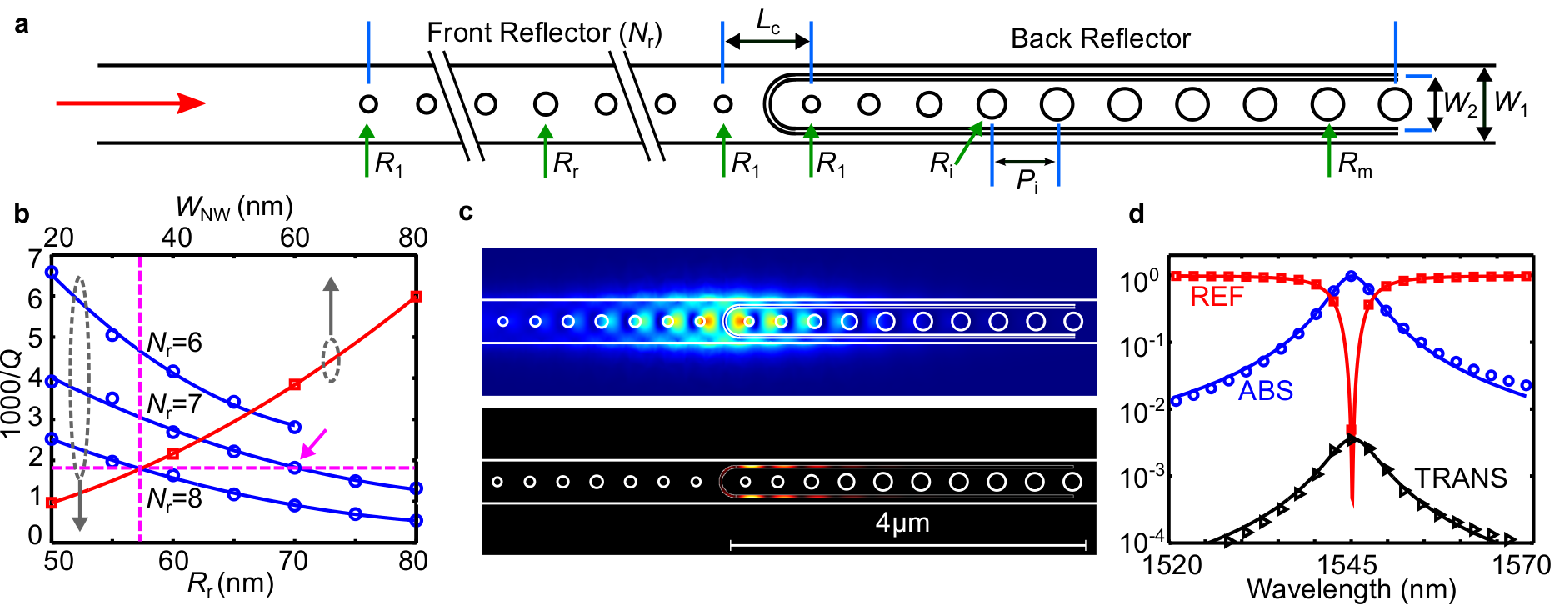}
\caption{Design of an integrated CPA-SNSPD. \textmd{a) Model for an asymmetric nanobeam cavity cavity enclosing a nanowire. Fixed parameters are: silicon waveguide width ($W_1=500$nm), nanowire layout ($W_2=350$nm), design of the back reflector (maximum hole radius $R_{\text{m}}=100$nm), minimum hole radius ($R_{\text{1}}=50nm$) and cavity length ($L_{\text{c}}=560$nm). Variable parameters are: number of holes ($N_{\text{r}}$) and maximum hole radius ($R_{\text{r}}$) of the front reflector, and nanowire width $W_{\text{NW}}$. For the pitch, $P_{\text{i}}=1.2256R_{\text{i}}+308$nm is used to center the photonic bandgap to 1545nm. The tapers are linear. b) Simulated $Q_{\text{A}}(W_{\text{NW}})$ (red squares), and $Q_{\text{r}}(R_{\text{r}},N_{\text{r}})$ (blue circles). The solid lines are the best polynomial fits. The dashed lines and pink arrow mark the design of CPA-SNSPDs presented in this paper. c) Simulated magnitude of the electric field on the waveguide surface (top) and power dissipation density in the nanowire (bottom) for a CPA-SNSPD ($W_{\text{NW}}=35$nm, $N_{\text{r}}=7$, $R_{\text{r}}=70$nm) excited with TE-polarized guided light from the left. d) Simulated absorption (ABS - blue circles), reflection (REF - red squares) and transmission (TRANS - black triangles) for the same device. The lines are Lorentzian fits. }}
\label{fig:2}
\end{figure}

In order for a short nanowire to absorb virtually every photon incident along a waveguide,  it is placed within an asymmetric optical microcavity defined by etching a series of holes on either side of the nanowire (see Fig.~\ref{fig:2}a). The cavity containing the weakly absorbing nanowire is designed to function as a coherent perfect absorber\cite{Wan2011,Chong2010,Zhang2012} (CPA).   Ideally (ignoring all other losses), the reflectivity of the back mirror (right side) is unity, so all of the light incident from the left must be either absorbed or reflected. At the resonant frequency of the cavity, $\omega_{\text{R}}$, a portion of the light incident from the left excites a resonating cavity mode with amplitude $A$. The normalized power reflected back into the incident waveguide is then $|1-\sqrt{2/\tau_{\text{r}}}A|^2$, and the power absorbed by the nanowire is $2|A|^{2}/\tau_{\text{A}}$, where $\tau_{\text{r}}$ and $\tau_{\text{A}}$ are the time constants associated with the decay of $A$ into the waveguide and the nanowire, respectively\cite{Joannopoulos2011}. Power conservation in this idealized scenario requires $|1-\sqrt{2/\tau_{\text{r}}}A|^{2}+2|A|^{2}/\tau_{\text{A}}=1$, which in general determines how much of the incident light reflects, and how much is transferred into the cavity. However, if the left side mirror reflectivity is designed such that $\tau_{\text{r}}=\tau_{\text{A}}$, then all of the incident light is perfectly transferred to the cavity and absorbed. From a different perspective, the scattering matrix ($M$) describing the dielectric mirrors and nanowire in a CPA-SNSPD is such that the incident photons excite the detector into an eigenvector of $M$ with eigenvalue equal to zero \cite{Wan2011}.

In the ideal scenario above, an arbitrarily short nanowire (arbitrarily long $\tau_{\text{A}}$) can absorb all of the incident radiation as long as the reflectivity of the input mirror can be precisely tuned close to 100\% (i.e. long $\tau_{\text{r}}$). In practice there will also be some additional scattering losses associated with the cavity, so its overall quality factor, $Q$ will be given by $Q^{-1}=Q_{\text{A}}^{-1}+Q_{\text{r}}^{-1}+Q_{\text{scatt}}^{-1}$, where $Q_{\text{A}}=\omega_{\text{R}}\tau_{\text{A}}/2$, and $Q_{\text{r}}=\omega_{\text{R}}\tau_{\text{r}}/2$. The above analysis indicates that detectors with absorption efficiency $\eta_{\text{A}}\sim 1$, can be achieved using very short and narrow nanowires as long as  dielectric reflectors can be designed such that $Q_{\text{A}}=Q_{\text{r}}\ll Q_{\text{scatt}}$. The minimum nanowire length ($L_{\text{NW}}$) is dictated by how tightly localized the high $Q_{\text{scatt}}$ cavity mode is, and therefore high-index contrast host materials like SOI are ideal. Finally it is noted that this approach necessarily limits the bandwidth over which the detector absorbs efficiently, to $\sim \omega_{\text{R}}/Q$.  However, for many quantum information processing applications, optical filters are placed in front of detectors to minimize spurious counts due to stray photons\cite{Yang2014,Zinoni2006}.  The CPA detector described here actually integrates the filter and detector into a single, compact unit.

\subsection{Design Details}
 The CPA-SNSPDs are formed by pattering holes in a silicon nanobeam (silicon on SiO$_2$, 190nm thick, 500nm wide) to form an asymmetric cavity around a U-shaped nanowire (NbTiN, 8nm thick, $W_{\text{NW}}$ wide) that lies on top of the nanobeam, as illustrated in Fig.~\ref{fig:2}a. Nanobeam cavities\cite{Deotare2009,Joannopoulos2011} allow low-loss coupling to waveguides, high $Q_{\text{scatt}}$, low mode volume, and an accessible near field which are all useful features for CPA-SNSPDs. The back (perfect) reflector consists of 10 holes, 6 of fixed radius ($R_\text{m}=100\text{nm}$) and 4 with shrinking radii, down to $R_1=50\text{nm}$, toward the input waveguide to impedance match the Bloch mode to the waveguide mode. The front (partial) reflector has $N_{\text{r}}$ holes, all of which are linearly tapered from a maximum hole radius of $R_{\text{r}}$ down to $R_1=50\text{nm}$ on both sides. These 1D photonic crystals\cite{Joannopoulos2011} were designed (see simulation methods) to have a band-gap centered at 1545nm. To design the CPA, $N_{\text{r}}=14$ and $R_{\text{r}}=100$nm are fixed, and the cavity length ($L_{\text{c}}$) in the absence of the nanowire is adjusted to get a mode at the detector operation wavelength of 1545nm with a moderate $Q=5.6\times10^4$. This ensures appropriate design of tapers and consequently a high enough $Q_{\text{scatt}}>5.6\times10^4$. When the nanowire is added to the same cavity, the mode profile and wavelength stay almost fixed, but the new $Q\simeq Q_{\text{A}}(W_{\text{NW}})$ is substantially reduced, even for the smallest $W_{\text{NW}}=20\text{nm}$ (see squares on Fig.~\ref{fig:2}b). Removing the nanowire and reducing $R_{\text{r}}$ and $N_{\text{r}}$ results in a new set of $Q\simeq Q_{\text{r}}(R_{\text{r}},N_{\text{r}})$, with magnitudes comparable with $Q_{\text{A}}(W_{\text{NW}})$ (see circles on Fig.~\ref{fig:2}b). A nanowire with a certain $W_{\text{NW}}$ (e.g. vertical dashed line on Fig.~\ref{fig:2}b) and thus fixed $Q_{\text{A}}$ (horizontal dashed line) will perfectly absorb incident coherent light if $N_{\text{r}}$ and $R_{\text{r}}$ are chosen such that $Q_{\text{r}}(R_{\text{r}},N_{\text{r}})=Q_{\text{A}}(W_{\text{NW}})$ (pink arrow on the same figure). Note that in this design algorithm, $L_{\text{NW}}$ is fixed by what is required for electrical connection through the back reflector ($\sim2\times$4\(\upmu\)m); this can in principle be further reduced by employing a better connection topology.

The fabricated CPA detectors have nominal parameters; $W_{\text{NW}}=35$nm, $N_{\text{r}}=7$, $R_{\text{r}}=70$nm, $Q_{\text{CPA}}\simeq (Q_{\text{A}}^{-1}+Q_{\text{r}}^{-1})^{-1}=275$, and $L_{\text{NW}}=8.5$\(\upmu\)m. The electric field and power loss distributions are as shown in Fig.~\ref{fig:2}c, for the structure excited with the mode of the strip waveguide at 1545nm. The field is tightly localized by the cavity, and the absorption is concentrated over a micron-length portion of the nanowire. At resonance, the calculated absorption peaks at 98.4\%, while reflection, transmission and radiated losses are 0.04\%, 0.3\%, and 1.1\% respectively (see simulated spectrum shown on Fig.~\ref{fig:2}d). Analyzing the variation of peak absorption with different fabrication imperfections (see supplementary info) shows the structure is quite robust. Analysis of the radiation losses shows that 1\% of the total 1.1\% is lost to the substrate. $Q_{\text{scatt}}$ could be further increased by undercutting the nanobeam, and the back-transmission could be further reduced by adding more holes. Applying both strategies, the losses would be reduced to $\sim$0.2\% and the absorption could peak close to $\sim$99.8\%. For comparison, the simulated absorption of the same 8.5\(\upmu\)m long nanowire, but without the cavity, is 29\%; a 160\(\upmu\)m long nanowire would be needed to obtain 99.8\% absorption in the absence of the cavity. The huge reduction of $L_{\text{NW}}$ without compromising absorption efficiency ($\eta_{\text{A}}$) promises a detector not only efficient in absorption, but also superior in the efficiency of converting absorbed photons to electric pulses ($\eta_{\text{D}}$). It also promises less dark counts, higher speed, and a more compact foot-print, as demonstrated and discussed below.

\subsection{Measurement Setup}
At the center of Fig.~\ref{fig:1} is a micrograph showing a portion of one of the fabricated chips. It consists of detectors with input waveguides (WG) connected to focusing grating couplers\cite{Vermeulen2012} (GC) as well as gold electrical connections that are routed to large contact pads (not shown). Also included on the chip are devices with two GCs connected by a waveguide; these were used for alignment and calibration purposes. The chip was installed on, and wire-bonded to, a custom-made printed circuit board that hosts electronic circuits to apply bias, preamplify the detection signal and impedance match the detector to a $50\Omega$ coaxial cable that runs to room-temperature amplifiers and then to a counter (see schematics on Fig.~\ref{fig:1}). The board was installed in an optical cryostat (SVT-300, Janis Inc.) where it was submerged in superfluid helium (T=2.05K). The whole sample holder is surrounded by a cylindrical brass shield with two holes (S on Fig.~\ref{fig:1}) of 6mm diameter coaxial with the two 45 degree windows (W on Fig.~\ref{fig:1}), one for the excitation laser light, and the other to image the chip for aligning and calibration. An attenuated and polarization-controlled continuous-wave (CW) tunable laser (TLB-6600, New Focus Inc.) was coupled to waveguides by focusing it on the input grating couplers. The setup was calibrated to read the rate of photons incident through the input cryostat window ($R_{\text{Ph}}$ on Fig.~\ref{fig:1}) by a power meter ($\text{D}_1$). Several measurements (see methods) were done to calibrate the coupling efficiency of the incident photons, $R_{\text{Ph}}$, into the strip waveguides ($\eta_{\text{C}}$), as needed for on-chip efficiency measurements ($\eta_{\text{C}}$ is shown by circles on Fig.~\ref{fig:3}a).

\subsection{Efficiency and Dark Count Characterization}
To measure the quantum efficiency and noise performance, three different count rates were determined (see methods): photon count rate (PCR) measured when the input grating couplers were excited by the laser, background (mostly black-body) count rate (BCR) measured in the same conditions but with the laser off, and  dark count rate (DCR) measured when a cylindrical brass shield (50\(\upmu\)m thick) without any holes surrounded the sample holder. The BCR will exceed the DCR because the large cryostat windows - that were unshielded for PCR and BCR measurements - allow intense blackbody radiation to enter the cryostat. The quantum efficiency ($\text{QE}=\eta_{\text{A}}\eta_{\text{D}}$) was deduced from the measured rates as $(\text{PCR}-\text{BCR})/(\eta_{\text{C}}R_{\text{Ph}})$.

\begin{figure}[h!]
\centering
\includegraphics[width=6.15in]{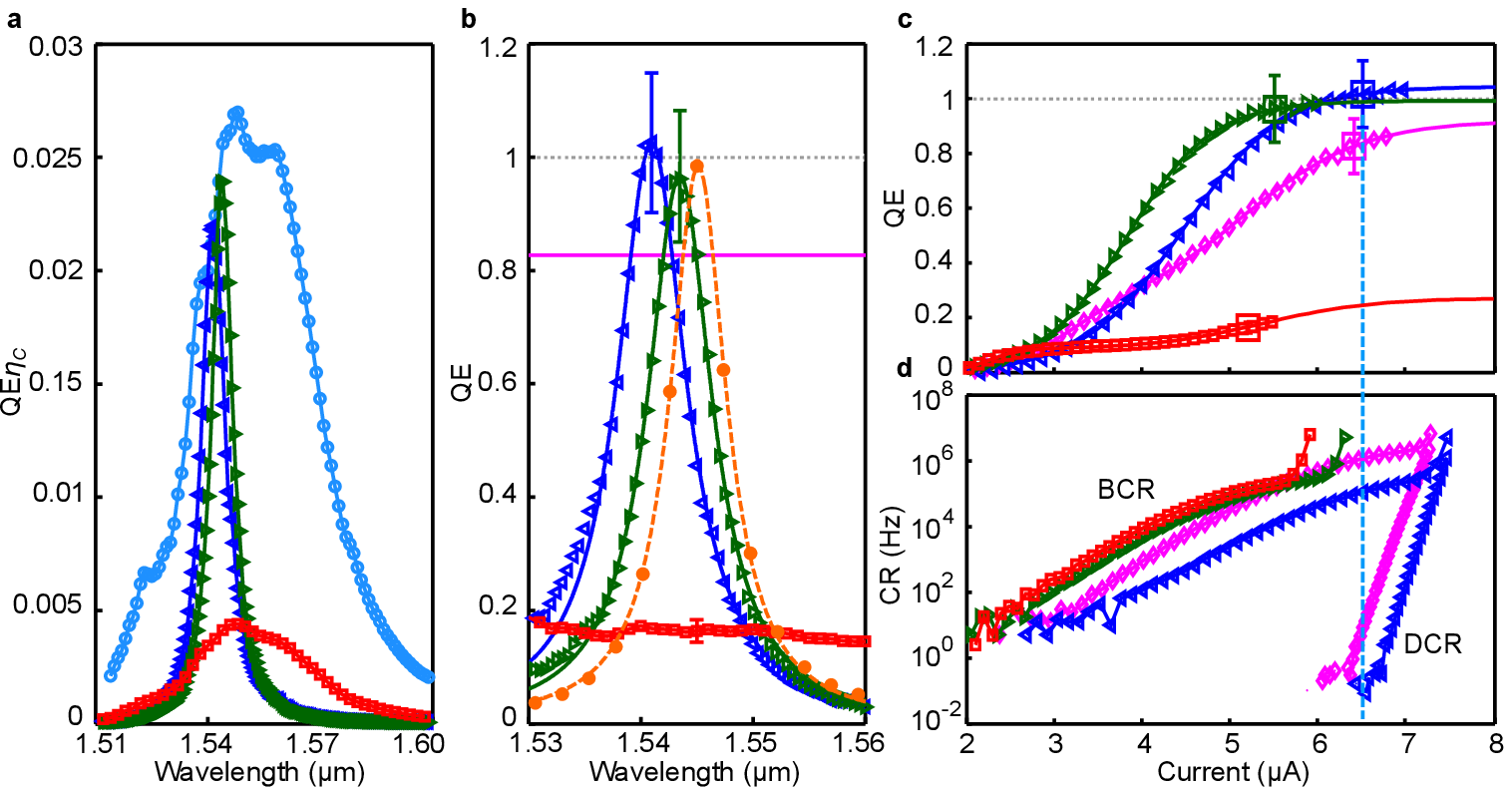}
\caption{Integrated SNSPD detection performance. \textmd{$\triangleleft$ and $\triangleright$ are for CPA based detectors; $\Box$ and $\diamond$ are for TW based detectors with $L_{\text{NW}}$ equal to 8.5\(\upmu\)m and 57.2\(\upmu\)m, respectively. Each symbol is for a fixed device. a) QE$\eta_{\text{C}}$ as well as $\eta_{\text{C}}$ (circles) measured over a wide bandwidth for detectors biased at $\sim$0.9$I_{\text{C}}$. b) On-chip measured QE and designed QE (filled circles and dashed line) for CPA SNSPDs measured versus wavelength at $\sim$0.9$I_{\text{C}}$. The solid lines drawn on the QE of CPA devices are the best fits to a the expected Lorentzian lineshape. Filled circles and a dashed line are for the simulated $\eta_{A}$. c) QE versus bias current measured at the cavity resonance wavelength for CPA-SNPSDs ($\triangleleft$ and $\triangleright$), and at 1550nm for TW SNSPDs ($\Box$ and $\diamond$). For each device, $\sim$0.9$I_{\text{C}}$ is marked by a square. d) DCR and BCR measured versus bias current. The vertical dashed line marks $\sim$0.9$I_{\text{C}}$ for a CPA-SNSPD and is merely for guiding the eye. All the detector measurements were done at 2.05K.}}
\label{fig:3}
\end{figure}

Figure~\ref{fig:3}a shows the system quantum efficiency, $\text{QE}\eta_{\text{C}}$, for several SNSPDs, all biased at a bias current equal to $\sim$90\% of the experimentally determined critical current ($I_{\text{C}}$). Triangles are for two CPA-SNSPDs, while squares are for a device with the same nanowire layout, but without any cavity holes, effectively a TW structure. The TW device exhibits a broad spectral response which is a down scaled version of $\eta_{\text{C}}$ (expected because of the small $\eta_{\text{A}}$ for such short TW SNSPD). In contrast, the CPA devices show resonant-like spectra that almost ideally sample the $\eta_{\text{C}}$. Note that the off-resonant $\text{QE}\eta_{\text{C}}$ goes to negligibly small values compared to the peak (only 0.3\% of the peak $\text{QE}\eta_{\text{C}}$) confirming negligible contribution of non-guided photons to the peak PCR.  

Although the efficiency numbers reported on Fig.~\ref{fig:3}a are small, a relevant number for integrated quantum optics applications is the efficiency of detecting photons that are already in the waveguide. With the same symbol conventions, Fig.~\ref{fig:3}b shows the on-chip QEs at $\sim0.9I_{\text{C}}$ obtained by normalizing the above measured $\text{QE}\eta_{\text{C}}$ with $\eta_{\text{C}}$. The $\text{QE}$ for devices without the cavity stays flat and small, as both $\eta_{\text{A}}$ and $\eta_{\text{D}}$ are broadband\cite{Marsili2012}, and as $\eta_{\text{A}}$ is expected to be much less than unity. The QE for the CPA devices peak close to unity (equal to unity within the uncertainty of the measurements) signaling very efficient $\eta_{\text{A}}$ and appropriate functioning of the CPA design, as well as very high $\eta_{\text{D}}$. Shown on the same figure, with filled circles and a dashed line (best Lorentzian fit), is the simulated $\text{QE}=\eta_{\text{A}}$ (assuming $\eta_{\text{D}}=1$) for the designed CPA detector. As can be seen the fabricated CPA devices perform close to the target QE. The small difference in the resonant wavelength is due to fabrication imperfections, and could be improved by using better fabrication tools.

In addition to wavelength, the quantum efficiency of a CPA SNSPD is a function of bias current. Figure~\ref{fig:3}c shows the QE versus bias current at a fixed wavelength for the same devices with the same symbols. The lines are best fits to experimental data using a double sigmoidal function. A typical bias current of $\sim$0.9$I_{\text{C}}$ for each device is marked with a big square. The QE of the two CPA devices (triangles) measured at their resonant frequencies follow a single sigmoidal shape with saturation at 1.04 and 0.99. These devices perform so well that the small difference between their measured saturated QE and unity stays less than our experimental uncertainties. However, their QEs at $\sim$0.9$I_{\text{C}}$ are $\sim$2\% smaller than the saturation level, showing a high $\eta_{\text{D}}$ of $\sim$98\%. This value together with the simulated value of $\eta_{\text{A}}=$98.4\% implies an upper bound for $\text{QE}=\eta_{\text{A}}\eta_{\text{D}}\sim$96\% at $\sim$0.9$I_{\text{C}}$. The QE for the non-cavity device (squares) measured at 1550nm follows a more complex double sigmoidal curve, signaling the presence of a material or geometrical constriction in the nanowire (in agreement with its smaller $I_{\text{C}}$). But, the fitted sigmoid has a saturation at $\text{QE}=\eta_{\text{A}}=$27\% ($\eta_{\text{D}}=$1 at saturation), close to the simulated $\eta_{\text{A}}$ of 29\%.

For comparison, a TW device with a simulated $\eta_{\text{A}}$ of 90\% was also fabricated using a relatively long nanowire ($W_{\text{NW}}=$35nm) length of 57.2\(\upmu\)m. The QE of this device (diamonds on Fig.~\ref{fig:3}c) changes considerably more gradually with the bias current than the CPA structures, most probably because of more non-uniformities along its long absorbing length. At $\sim$0.9$I_{\text{C}}$ the QE  is only 89\% of its saturated level at $\sim$92\%. This then clearly demonstrates the advantage of reducing $L_{\text{NW}}$ in CPA detectors, which allows high $\eta_{\text{D}}$ and high $\eta_{\text{A}}$, therefore a high QE.

Quite apart from being efficient, a good detector must be low noise. Figure~\ref{fig:3}d shows DCR and BCR measurements for devices of Fig.~\ref{fig:3}c with the same symbols. The DCR starts from a plateau-like level at around 0.1Hz followed by a fast exponential increase as expected for nanowire detectors. The plateau likely originates from black-body radiation that enters the cryostat from top or even through the thin brass foil; it can be reduced by using better shields\cite{Schuck2013a,Yang2014}. At $\sim$0.9$I_{\text{C}}$ the DCR for the non-cavity 57.2\(\upmu\)m long device is $\sim$1.1Hz, whereas the DCR for the CPA device is $\sim$0.1Hz. This confirms another advantage of the CPA design, in substantially reducing the DCR. Note that a CPA-SNSPD at $\sim$0.9$I_{\text{C}}$ (see vertical dashed line connecting figures~\ref{fig:3}c and d) maintains its high QE over a bias current range for which DCR stays negligibly small (i.e. almost ideal detection performance). The BCR versus bias current curves show QE-like shapes which further suggests it originates from black-body or stray photons rather than being intrinsic to the detectors, like the DCR. This is in agreement with other studies that show how the BCR can be diminished by using well-shielded fiber coupled cryostats and utilizing inline cold filters\cite{Schuck2013a,Yang2014}.

\subsection{Timing Performance Characterization}
\begin{figure}[h!]
\centering
\includegraphics[width=6.15in]{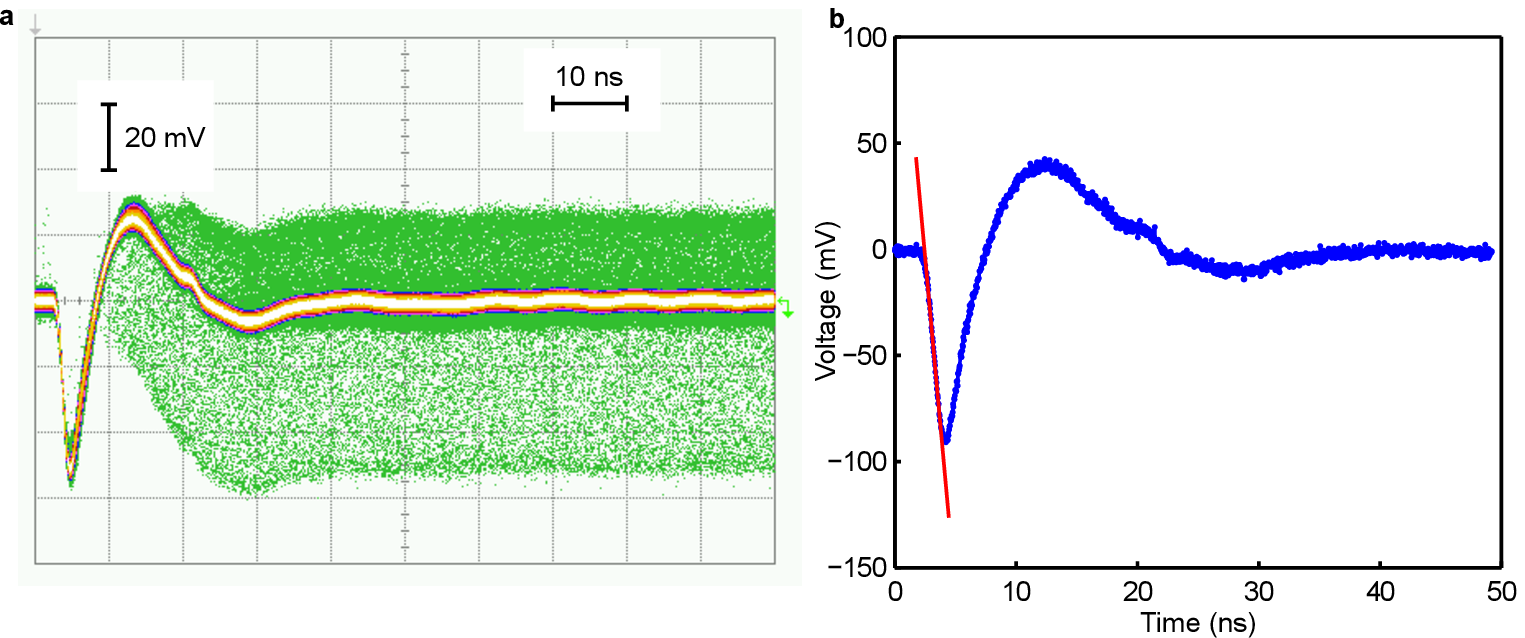}
\caption{Performance of CPA SNSPDs in time. \textmd{a) Waveform histogram of detection pulses when the detector was excited by a CW laser. b) A single photon detection trace with falling slope ($K$) marked by a red line.}}
\label{fig:4}
\end{figure}

Figure~\ref{fig:4}a shows a waveform histogram of amplified photon detection pulses from the detector when biased at 3\(\upmu\)A and excited by a CW laser. An inductor, $L=$100nH, was externally placed in series with the nanowire and the two were looking at an impedance of $R=100\Omega$ (see Fig.~\ref{fig:1}). The measured pulse has a negative polarity making it compatible with the counter, and shows an under-damped shape because of the frequency cut-offs of the chain of amplifiers (10MHz to 1.2GHz). The counts are restored about 7ns after an initial detection event at the falling edge, approximately five $L/R$ time constants, as expected for an $LR$ circuit. The external inductor was used in the measurement setup to ensure a smooth over-damped return of bias current to the nanowire, and therefore to avoid after pulses\cite{Burenkov2013}. However, this inductor can in principle be removed and replaced with a more careful read-out circuit to further reduce the reset time to its intrinsic limit, that is known to scale with nanowire length \cite{Kerman2009,Kerman2006,Akhlaghi2012}.

As for the timing jitter, the energy stored in a CPA cavity decays with a time constant $\tau=Q_{\text{CPA}}/\omega_{\text{R}}$. This gives $\tau=$0.23ps for the CPA-SNSPD designs reported here. Provided $\tau$ is much smaller than the minimum reported timing jitter for nanowire detectors ($<$15ps), this cavity decay time constant should not have any adverse effect on the measured timing jitter. As an estimation of the timing jitter of this system, a single photon detection trace at 5\(\upmu\)A is shown in Fig.~\ref{fig:4}b.  The associated root-mean-square of the electrical noise is $\sigma_{\text{n}}=$1.4mV, and the slope of falling edge is $K=$62.7mV/ns. This yields an estimated timing jitter of $2.355\sigma_{\text{n}}/K=$53ps full-width at half-maximum.  This jitter is limited by the performance of the read-out circuits, and is consistent with reports on SNSPDs with comparable bias currents\cite{MarsiliF.2013}.

\section{Conclusion and Outlook}
The demonstrated impact of coherent perfect absorption on the performance of a single photon detector paves the way for integrating hundreds of ultra-high performance detectors on a chip. The silicon host material facilitates high-performance CPA-SNSPDs because of its high index contrast, while also offering the potential of integrating on-chip electronic circuits with the detectors to further boost the performance, and to handle the complexity of scaling up the read-out circuits. This latter point is especially interesting in light of studies that show the compatibility of CMOS transistors with the cryogenic environment\cite{Das2011,Tack1989}. The projected superior speed performance of CPA based detectors combined with their built-in filtered QE, and advanced fiber to waveguide coupling methods\cite{Roelkens2006}, will also make these detectors ideal for implementing fast quantum communication systems.

\section{Methods}

\subsection{Simulations}
Frequency domain finite element solvers (COMSOL, Inc.) are used for numerical simulations. The index of refraction for silicon is set to 3.45 (measured close to 1550nm at cryogenic temperatures\cite{Frey2006}), and the environment is assumed to be superfluid helium ($n=$1.03). The index of refraction for NbTiN is set to 4.17+i5.62\cite{Tanner2010}. Eigenmode analysis is used for all $Q$ factor simulations, while the simulated spectra are obtained by using available numeric port boundary conditions. 

\subsection{Fabrication}
The devices were fabricated on NbTiN coated (8nm thick - STAR Cryoelectronics Inc.) SOI wafers with a silicon device layer thickness of 200nm and a buried oxide thickness of 1220nm. The superconducting thin film has a critical temperature $T_{\text{C}}=$7.16K, and critical current density $J_{\text{C}}(T=0)=7.57\times10^{6}$A/cm$^{2}$ (see supplementary information for details). Positive electron-beam (e-beam) resist (ZEP520A from ZEON Corp.) was spin-coated at 1800rpm and hot-plate backed at 180$^{\circ}$C for 3min to make a 600nm thick film. Contact pads and the first set of alignments marks were defined in a 25KeV e-beam lithography machine (dose 110\(\upmu\)C/cm$^{2}$) and developed in o-xylene at 20$^{\circ}$C for 60s followed by a 30s soak in IPA and a DI water rinse. The chip was rinsed in 140:1 BHF:H$_{2}$O for 60s to wet-etch $\sim$1nm of NbTiN surface oxides, and was immediately  transferred to an e-beam evaporator chamber to deposit an 8nm/90nm titanium-gold bilayer, followed by lift-off in sonicated chlorobenzene. The chip was coated again by 600nm thick ZEP520A, after which the photonic structures and a second set of alignments marks were e-beam written using the original gold alignment marks at 140\(\upmu\)C/cm$^{2}$.  The sample was then developed in cold o-xylene (4$^{\circ}$C - to improve contrast\cite{Yang2007}). Reactive ion etching (RIE) in 15:2 CF$_{4}$:O$_{2}$ 30s and 15:1 CF$_{4}$:O$_{2}$ $\sim$300s was used to vertically etch through the entire thickness of unprotected NbTiN and silicon to yield the NbTiN coated photonic structures.  Spin-coating of 1:1 Anosole:ZEP520A at 2800rpm and the same baking conditions were used to coat 180nm of resist on flat areas and 110nm on waveguides that are surrounded by relatively deep trenches. The second alignments marks were uncovered by e-beam lithography and development to allow sharp imaging of the marks for the last lithography step. The nanowires were written using these uncovered marks at 78mC/cm$^{2}$ for which ZEP520A acts as a negative resist\cite{Oyama2011}. All of the resist except the area exposed by the high-dose e-beam was removed by 5min exposure to ultraviolet radiation ($\lambda=320\text{nm}$, $\sim3\text{W/cm}^2$) and a 1min rinse in chlorobenzene. A final RIE in 15:2 CF$_{4}$:O$_{2}$ for $\sim$40s was used to etch the unprotected NbTiN and 10nm of silicon. SEM images of several samples indicates this process yields better than $\pm$20nm alignment of the nanowires with the nanobeam cavities.

\subsection{Measuring $\eta_{\text{C}}$}
 It is very difficult to {\it directly} measure $\eta_{\text{C}}$ - the coupling efficiency between the photons incident on the cryostat windows at room temperature and the strip waveguide - as there is no direct means of accessing the strip waveguide inside the cryostat.  The $\eta_{\text{C}}$ can however be {\it indirectly} but accurately inferred by incorporating several test devices on the chip, as follows.

The first test device is laid out as two grating couplers connected with a long waveguide (see Fig.~\ref{fig:1}) that has a U-shaped nanowire on top (the same nanowire geometry as the TW SNSPD devices presented, but without contacts.). Seven of these devices with nanowire lengths ($L_{\text{NW}}$) from 0 to 60\(\upmu\)m were made and their transmission ($T$) measured.  A linear fit to log$(T/T(L_{\text{NW}}=0))$ versus $L_{\text{NW}}$ yields $\eta_{\text{A}}$ versus $L_{\text{NW}}$ for devices without cavities. At $L_{\text{NW}}=$57.2\(\upmu\)m the measured $\eta_{\text{A}}=$92.8\%, and the best linear fit gives 91.7\%, both in good agreement with the designed $\eta_{\text{A}}$ of 90\%. The QE$\eta_{\text{C}}$=$(\text{PCR}-\text{BCR})/R_{\text{Ph}}$ at 1550nm for the 57.2\(\upmu\)m long SNSPD versus bias current is then measured (scaled version of diamonds on Fig.~\ref{fig:3}c). Using the best sigmoidal fit to the measured points, the maximum saturated QE$\eta_{\text{C}}$ (for which $\eta_{\text{D}}$=1) and QE$\eta_{\text{C}}$ at $0.9I_{\text{C}}$ were determined and divided to yield $\eta_{\text{D}}=$89\% at $0.9I_{\text{C}}$. Then, measuring QE$\eta_{\text{C}}$ versus wavelength at $0.9I_{\text{C}}$ and equating the QE at 1550nm to $\eta_{\text{A}}\eta_{\text{D}}=$0.917$\times$0.89 (solid horizontal line on Fig.~\ref{fig:3}b), $\eta_{\text{C}}$ is calculated as shown by the circles on Fig.~\ref{fig:3}a.

A nice feature of the above procedure is that the uncertainties in absolute calibration of the power meter used to measure $R_{\text{Ph}}$ - the rate of photons incident to the cryostat windows - is mostly embedded in $\eta_{\text{C}}$ rather than QE. This keeps the estimated QE values largely independent of the meter's calibration. However, there may be variations in $\eta_{\text{C}}$ among different devices on the chip, causing errors in measured QE. To evaluate these variations, 14 identical test devices was included on the chip, each laid out as two grating couplers connected by a long waveguide. Measuring transmission through these devices, $\pm$12\% of change relative to the mean value is observed for wavelengths close to 1550nm. Noting the QE measurements like the transmission measurements involving two couplers, and considering that the above measurement of $\eta_{\text{A}}$ for $L_{\text{NW}}$=57.2\(\upmu\)m yields an $\eta_{\text{A}}$ very close to that expected, the uncertainty bars on QE as shown on Fig.~\ref{fig:3} were set at $\pm$12\%.

\subsection{Measuring Count Rates}
The count rates (CR) are by definition the number of counts registered by a counter per unit time during which the detector and all the readout electronics are active. Because the counter used in the setup (PicoHarp 300, PicoQuant Inc.), has a dead time ($T_{\text{DT}}$) of 87ns, in all of the CR measurements reported here, the effect of $T_{\text{DT}}$ was included by defining CR = CR$_{\text{m}}$/(1- CR$_{\text{m}}T_{\text{DT}}$), where CR$_{\text{m}}$ is simply the number of counts divided by counting time. The linearity of CR with $R_{\text{Ph}}$ confirms that the measured CRs are neither affected by $T_{\text{DT}}$ or by detector nonlinearities.

\section{Acknowledgements}
Financial support from the Canadian Institute for Advanced Research and the Natural Sciences and Engineering Research Council is gratefully acknowledged.

\section{References}

\end{document}